\documentclass[a4paper,fleqn,usenatbib]{mnras}

\usepackage{txfonts,pdflscape,longtable}

\usepackage[T1]{fontenc}
\usepackage{ae,aecompl}

\usepackage{graphicx}	% Including figure files
\usepackage{amssymb}	% Extra maths symbols
\usepackage{float}
\usepackage{color}

\title[New mass-loss recipe]{Mass loss and the Eddington parameter: a new mass-loss recipe for hot and massive stars}

\author[J. M. Bestenlehner]{
Joachim\,M.\,Bestenlehner$^{1}$\thanks{E-mail: j.m.bestenlehner@sheffield.ac.uk}
\\
$^{1}$Department of Physics \& Astronomy, Hounsfield Road, University of Sheffield, S3 7RH, UK\\
}

\date{Accepted XXX. Received YYY; in original form ZZZ}

\pubyear{2020}

\begin{document}
\label{firstpage}
\pagerange{\pageref{firstpage}--\pageref{lastpage}}
\maketitle

\begin{abstract}
Mass loss through stellar winds plays a dominant role in the evolution of massive stars. In particular the mass-loss rates of very massive stars (VMSs, $> 100\,M_{\odot}$) are highly uncertain. Such stars display Wolf-Rayet spectral morphologies (WNh) whilst on the main-sequence. Metal-poor VMSs are progenitors of gamma-ray bursts and pair instability supernovae. In this study we extended the widely used stellar wind theory by Castor, Abbott \& Klein from the optically thin (O star) to the optically thick main-sequence (WNh) wind regime. In particular we modify the mass-loss rate formula in a way that we are able to explain the empirical mass-loss dependence on the Eddington parameter ($\Gamma_{\rm e}$). The new mass-loss recipe is suitable for incorporation into current stellar evolution models for massive and very massive stars. It makes verifiable predictions, namely how the mass-loss rate scales with metallicity and at which Eddington parameter the transition from optically thin O star to optically thick WNh star winds occurs. 
In the case of the star cluster R136 in the Large Magellanic Cloud we find in the optically thin wind regime $\dot{M} \propto \Gamma_{\rm e}^{3}$ while in the optically thick wind regime $\dot{M} \propto 1/ (1 - \Gamma_{\rm e})^{3.5}$. The transition from optically thin to optically thick winds occurs at $\Gamma_{\rm e, trans} \approx 0.47$. The transition mass-loss rate is $\log \dot{M}~(M_{\odot} \mathrm{yr}^{-1}) \approx -4.76 \pm 0.18$, which is in line with the prediction by Vink \& Gr\"afener assuming a volume filling factor of $f_{\rm V} = 0.23_{-0.15}^{+0.40}$.
\end{abstract}

\begin{keywords}
stars: Wolf-Rayet -- stars: early-type -- stars: atmospheres -- stars: mass-loss -- stars: winds, outflows
\end{keywords}

%%%%%%%%%%%%%%%%%%%%%%%%%%%%%%%%%%%%%%%%%%%%%%%%%%

%%%%%%%%%%%%%%%%% BODY OF PAPER %%%%%%%%%%%%%%%%%%

\section{Introduction}
The physics and evolution of massive stars remain unclear owing to uncertainties in nuclear reaction rates, stellar structure, internal mixing processes and especially mass-loss properties \citep{langer2012}. Mass loss plays a key role during the evolution of massive stars and determines the final stellar mass before ending their life as core-collapse supernova \citep[e.g.][]{heger2003} and/or potentially as long duration gamma-ray burst \citep[LGRBs,][]{woosley2006b}. Hot, massive stars lose mass through radiation driven stellar winds, which removes angular momentum from stars. The angular momentum loss influence the rotation properties and evolutionary path of massive stars \citep[e.g.][]{langer1998, meynet2000, brott2011} and their potential end as a LGRB \citep{woosley2005, woosley2006a}.

The widely used radiation driven wind theory has been developed in the 70s by \citet*[CAK hereafter]{castor1975}. CAK and its extensions and modifications are able to successfully reproduce the fundamental properties of OB star stellar winds \citep[e.g.][]{friend1986, pauldrach1986}. Solving the equation of motion in the single scattering limit has led to mass-loss predictions for O stars \citep{abbott1982, pauldrach1986, kudritzki1989}. These mass-loss predictions are typically lower than observed. \cite{puls1996} suggested that the discrepancy can be resolved by introducing a multi-scattering approach. Monte-Carlo line-transfer models have been used to estimate the line force including multiple scattering events which has led to mass-loss predictions \citep[e.g.][]{pauldrach2001, vink2000, vink2001}. The mass-loss recipes by \cite{vink2000, vink2001} are usually used in stellar structure calculations for massive main-sequence stars while mainly empirical mass-loss recipes such as \cite{nugis2000} are used for Wolf-Rayet stars.

The mass loss through stellar winds strongly depends on the Eddington parameter $\Gamma_{\rm e}$ \citep{vink2002, vink2006, graefener2008, graefener2011}. It steeply increases at the transition from optically thin O star to optically thick Of/WN and  WNh star winds, which has been theoretically predicted by \cite{vink2011} and observationally confirmed by \cite{bestenlehner2014}. These very massive stars (VMSs, $> 100\,M_{\odot}$ \cite{vink2015:IAU}) display Wolf-Rayet spectral morphologies (WNh) whilst on the main-sequence. In the optically thin wind regime the mass-loss rates ($\dot{M}$) agree reasonably well with CAK while largely disagree in the optically thick wind regime \citep[e.g.][]{bestenlehner2014}. One reason might be the modest $1/(1 - \Gamma_{\rm e})^{\sim 0.7}$ term in CAK ($\alpha \approx 0.6$), which only boosts a steep increase in mass-loss at $\Gamma_{\rm e}$ close to unity
\begin{equation}\label{e:cak}
\dot{M}\propto M \frac{\Gamma_{\rm e}^{1/\alpha}}{(1-\Gamma_{\rm e})^{(1-\alpha)/\alpha}}
\end{equation}
with CAK fore multiplier parameter $\alpha$ and stellar mass $M$. Recent self-consistent stellar atmosphere models using full non-local thermal-equilibrium radiative transfer predict the velocity field and mass-loss rates of massive stars, but they are computational too expensive to be used on top of evolutionary stellar-structure calculations \citep{graefener2005, sundqvist2019, sander2020}. 

In this study we extend the CAK theory from optically thin to optically thick winds. We replace the stellar mass term in the CAK-description to account for the effect that the mass -- luminosity relation of massive stars becomes linear when approaching the Eddington limit ($\Gamma \rightarrow 1 \Rightarrow L \propto M$, e.g. \cite{yusof2013}). In this way we introduce an additional $\Gamma_{\rm e}^{1/2}$ and $1/(1-\Gamma_{\rm e})^2$ dependence, and resolve the discrepancy of CAK for optically thick winds. We test our relation on main-sequence O and hydrogen-burning Wolf-Rayet stars (type WNh) for the star cluster R136 in the Large Magellanic Cloud. A future study will focus on hydrogen free and evolved massive stars and test the updated stellar wind theory on {\it classical} Wolf-Rayet stars.

The current study is based on the original CAK wind theory, more specifically the mass-loss rate formula, and is structured as followed. In Sect.\,\ref{s:new-wind-theory} we derive our new mass-loss recipe by replacing the stellar mass term in CAK with a stellar mass -- Eddington parameter relation using the Eddington stellar model for radiative stars (Sect.\,\ref{s:edd-mass}) introducing a stronger dependence of the CAK wind theory on the Eddington parameter (Sect.\,\ref{s:new_cak}). In the discussion section (Sect.\,\ref{s:disc}) we test our updated CAK-type mass-loss recipe on observations and discuss its potential to predict mass-loss rates for all type of hot, massive stars. We conclude with a brief summary in Sect.\,\ref{s:con}.

\section{Mass-loss rates and the Eddington parameter: a new mass-loss recipe}
\label{s:new-wind-theory}
The mass-loss rate is the most important property for the evolution of the most massive stars. Stellar winds are parametrised via the mass-loss rate, terminal velocity, velocity law and wind inhomogeneity (clumping or volume filling factor). Theoretical and observational mass-loss rates show a strong dependence on the Eddington parameter \citep{graefener2008, vink2011, bestenlehner2014}. In the following section we take a closer look at the mass loss of the most massive stars and the dependence on the {\it classical} Eddington parameter considering only the electron scattering opacity ($\Gamma_{\rm e}$). In Sect.\,\ref{s:edd-mass} we introduce the Eddington stellar model and derive a scaling relation for the stellar mass with $\Gamma_{\rm e}$. Using this relation we replace the stellar mass term of the original CAK mass-loss rate formula and obtain a mass-loss recipe where $\dot{M}$ only depends on $\Gamma_{\rm e}$, the mean molecular weight ($\mu$) and the CAK force multiplier parameters (Sect.\,\ref{s:new_cak}). We discuss the validity of the Eddington stellar model for massive stars (Sect.\,\ref{s:valid}) and compare the $M-\Gamma_{\rm e}$ relation to stellar structure calculations (Sect.\,\ref{s:comp_mass}) and observations (Sect.\,\ref{s:comp_obs_mass}).

\subsection{The stellar model of Eddington and the stellar mass-Eddington parameter relation\label{s:edd-mass}}

The Eddington stellar model makes the following assumption about the star: (1) the energy transport is fully radiative, (2) the total pressure $P$ consist of the sum of gas pressure $P_{\rm gas}$ of a fully ionised ideal gas and the radiation pressure $P_{\rm rad}$ ($P = P_{\rm gas} + P_{\rm rad}$) and (3) the ratio of gas pressure to total pressure $P_{\rm gas}/P = \beta$ is constant throughout the star. 

In this case the energy transport through convection is neglected and the energy transport equation can be approximated by
\begin{equation}
\bigtriangledown\equiv \frac{\mathrm{d}\ln T}{\mathrm{d}\ln P} = \frac{1}{4}\frac{P}{P_{\rm rad}}\frac{\mathrm{d}P_{\rm rad}}{\mathrm{d}P}.
\end{equation}
We assume that the star is in a quasi-hydrostatic equilibrium 
\begin{equation}\label{e:hyd}
\frac{\mathrm{d}P}{\mathrm{d}r} = -\rho\frac{GM_r}{r^2}
\end{equation}
with the radius ($r$), density ($\rho$), the radius dependent mass ($M_r$) and the gravitational constant ($G$). The radiative acceleration can be expressed as
\begin{equation}\label{e:rad}
\frac{\mathrm{d}P_{\rm rad}}{\mathrm{d}r} = -\rho\frac{\kappa_r L_r}{4\pi c r^2}
\end{equation}
with the luminosity ($L_r$), the opacity ($\kappa_r$) by mass and the speed of light ($c$). Dividing Eq.\,\ref{e:rad} by Eq.\,\ref{e:hyd} we obtain
\begin{equation}\label{e:prad-p}
\frac{\mathrm{d}P_{\rm rad}}{\mathrm{d}P} = \frac{\kappa_r}{4\pi c G} \frac{L_r}{M_r}.
\end{equation}
Near the stellar surface, where the optical depth $\tau$ and the pressure approach zero ($P_0$, $P_{\rm rad,0}$), $M_r \approx M$ and $L_r \approx L$ and we find the following solution for Eq.\,\ref{e:prad-p}  
\begin{equation}\label{e:edd}
\frac{P_{\rm rad} - P_{\rm rad,0}}{P - P_0} \approx \frac{P_{\rm rad}}{P}= \frac{\kappa}{4\pi c G} \frac{L}{M} = (1-\beta) ~=~ \Gamma_{\rm e},
\end{equation}
where $\Gamma_{\rm e}$ is the {\it classical} Eddington parameter considering only the electron scattering opacity. In the Eddington stellar model only the ideal gas and radiation contribute to $P$. Therefore, the star is a polytrope with $n=3$ and
\begin{equation}\label{e:p_rho}
P = \left[\frac{3c}{4\sigma}\left(\frac{R}{\mu}\right)^4\frac{1-\beta}{\beta^4}\right]^{1/3} \rho^{4/3} = K\rho^{4/3},
\end{equation}
where $\sigma$ is the Stefan-Boltzmann-radiation constant, $R$ is the universal gas constant and $\mu^{-1} \approx 2X + 0.75Y + 0.5Z$ is the mean molecular weight with the chemical composition of hydrogen ($X$), Helium ($Y$) and metals ($Z$) in mass fraction. Using the Lane-Emden equation and the knowledge of a polytrope with $n=3$ the mass of the star is given as
\begin{equation}\label{e:le}
M = -\frac{1}{\sqrt{4\pi}}\left(\frac{4}{G}\right)^{3/2}K^{3/2} \xi_1^2\left(\frac{\mathrm{d}\theta}{\mathrm{d}\xi}\right)_{\xi=\xi_1},
\end{equation}
where $\xi_1^2\left({\mathrm{d}\theta}/{\mathrm{d}\xi}\right)_{\xi=\xi_1} \approx -2.01824$ is the Lane-Emden constant for a polytrope of $n=3$. Combining Eqs.\,\ref{e:edd}, \ref{e:p_rho} and \ref{e:le} we find an expression for the stellar mass
\begin{equation}\label{e:m-gamma}
M = \mathcal{C}\frac{1}{\mu^2}\frac{\Gamma_{\rm e}^{1/2}}{(1-\Gamma_{\rm e})^2},
\end{equation}
where $\mathcal{C}$\footnote{$\mathcal{C} = -\frac{2}{G^{3/2}}\left(\frac{3c}{\pi\sigma}\right)^{1/2}R^2\xi_1^2\left(\frac{\mathrm{d}\theta}{\mathrm{d}\xi}\right)_{\xi=\xi_1}$} includes all the constants from these equations. The stellar mass depends only on the Eddington parameter and the mean molecular weight ($\mu$) determined by the chemical composition. The $M-\Gamma_{\rm e}$ relation (Eddington mass) behaves as expected and $L\propto \mu^4M^3$ for $\Gamma_{\rm e}\ll1$ and $L\propto M$ for $\Gamma_{\rm e}\rightarrow1$ \citep{yusof2013}.

\subsection{Mass loss-Eddington parameter relation}
\label{s:new_cak}
\begin{table}
	\centering
	\caption{For given CAK force multiplier $\alpha$ we list expected transition Eddington parameters and mass-loss rate dependence for $\Gamma_{\rm e} \ll 1$ and $\Gamma_{\rm e} \rightarrow 1$.}
	\label{t:trans}
	\begin{tabular}{cccc}
		\hline
		$\alpha$ & $\Gamma_{\rm e,trans}$ & $\dot{M} \propto \Gamma_{\rm e}^{1/\alpha+1/2}$ & $\dot{M} \propto 1/(1 - \Gamma_{\rm e})^{(1-\alpha)/\alpha+2}$ \\
		\hline
		0.3 & 0.479 & $\Gamma_{\rm e}^{3.83}$ & $(1 - \Gamma_{\rm e})^{-4.3}$\\
		0.4 & 0.473 & $\Gamma_{\rm e}^{3.0}$ & $(1 - \Gamma_{\rm e})^{-3.5}$\\
		0.5 & 0.468 & $\Gamma_{\rm e}^{2.5}$ & $(1 - \Gamma_{\rm e})^{-3.0}$\\
		0.6 & 0.464 & $\Gamma_{\rm e}^{2.17}$ & $(1 - \Gamma_{\rm e})^{-2.7}$\\
		\hline
	\end{tabular}
\end{table}
In this section we combine the $M-\Gamma_{\rm e}$ relation (Eq.\,\ref{e:m-gamma}) with the standard CAK wind theory for massive stars. We used the original equation (46) from CAK (Eq.\,\ref{e:cak} for a simplified version) and then substituted the stellar mass with the $M-\Gamma_{\rm e}$ relation yielding to
\begin{equation}\label{e:cak_mdot-gamma}
\dot{M} = \mathcal{C}\frac{4 \pi G}{\kappa_{\rm e} \varv_{\rm th}}\frac{1}{\mu^2} k^{1/\alpha} \alpha(1-\alpha)^{(1-\alpha)/\alpha}\frac{\Gamma_{\rm e}^{1/\alpha+1/2}}{(1-\Gamma_{\rm e})^{(1-\alpha)/\alpha+2}},
\end{equation} 
where $k$ and $\alpha$ are the force multiplier parameters as defined in Eq. (12) of CAK, $\varv_{\rm th}$ is the thermal velocity and $\kappa_{\rm e}$ is the free electron opacity. The mass-loss rate only depends on the chemical composition (mean molecular weight), the {\it classical} Eddington parameter and the CAK force multiplier parameters, which are in some extent metallicity dependent (Table\,3 from \citet{puls2000}). A closer look at Eq.\,\ref{e:cak_mdot-gamma} shows, that there are two dependencies of $\dot{M}$. If $\Gamma_{\rm e} \ll 1$, $\dot{M} \propto \Gamma_{\rm e}^{1/\alpha+1/2}$. For $\Gamma_{\rm e} \rightarrow 1$, $\dot{M} \propto 1/(1 - \Gamma_{\rm e})^{(1-\alpha)/\alpha+2}$. Now we define the transition Eddington parameter ($\Gamma_{\rm e,trans}$), where the mass-loss dependency change from one relation to the other
\begin{equation}
\Gamma_{\rm e,trans}^{1/\alpha+1/2} = (1-\Gamma_{\rm e,trans})^{(1-\alpha)/\alpha+2}.
\end{equation}
As the solutions for such an equation are not straightforward and also imaginary solutions are possible we only list the real number solutions ($\Gamma_{\rm e,trans}$) for specific values of $\alpha$ in Table\,\ref{t:trans}. At lower metallicity $\alpha$ becomes smaller \citep[Table\,3,][]{puls2000} and $\Gamma_{\rm e,trans}$ moves to larger values. In addition, the slope below $\Gamma_{\rm e,trans}$ is steeper and $(1-\Gamma_{\rm e})$ dependence is stronger above. For O stars $\alpha \approx 0.6$ is a typical value while $\alpha$ is expected to be smaller at low metallicities \citep{puls2008}.

Replacing $M$ with the $M-\Gamma_{\rm e}$ relation adds an additional $\Gamma_{\rm e}^{1/2}/(1-\Gamma_{\rm e})^2$ dependence to CAK (Eq.\,\ref{e:cak_mdot-gamma}). The transition from $\dot{M} \propto \Gamma_{\rm e}^x$ and to the steeper $\dot{M} \propto 1/(1 - \Gamma_{\rm e})^y$ dependence occurs already for $\Gamma_{\rm e} \approx 0.5$ and not close to unity. A enhanced mass-loss rate at such a low $\Gamma_{\rm e}$ value is observed for Of/WN and WNh stars \citep{bestenlehner2014}.

\subsection{Validity of the Eddington stellar model\label{s:valid}}
In the Eddington stellar models the star is fully radiative. In the envelopes of hot, massive stars the energy transport is mainly radiative and convection can be neglected. For example, O stars have a convective core and probably a convective outer zone as well, but they have large radiative envelopes. The assumption that massive stars are radiative near the stellar surface is also adopted in stellar atmosphere calculations with radiation-driven winds which are used to analyse and study the physical properties OB and Wolf-Rayet stars, e.g. CMFGEN \citep{hillier1998}, FASTWIND \citep{puls2005} or PoWR \citep{Hamann2003}.
 
For stars with $T_{\rm eff}\geq 30$\,000\,K we can consider the gas to be fully ionised. The electron scattering opacity ($\kappa_{\rm e}$) is usually constant throughout the star and depends on the chemical composition of hydrogen and helium. In reality, some metals will not be fully ionised if the metallicity is not zero. The actual condition for hydrogen-rich main-sequence stars is $(1-\beta) \geq \Gamma_{\rm e}$ (Eq.\,\ref{e:edd}). The chemical compositions can only be determined at the stellar surface and introduces an additional bias if the star is not chemical homogeneous. This implies that the Eddington mass does not fully represent the true stellar mass. Potential consequences are discussed in the following section \ref{s:comp_mass}. In the case of evolved, hydrogen depleted {\it classical} WR stars the electron scattering opacity is less dominant near the stellar surface and $(1-\beta) >> \Gamma_{\rm e}$ can occur. This can lead to a significant underestimation of the true stellar mass.

With $\Gamma_{\rm e}$ considered to be constant throughout the star, $P_{\rm gas}$ and $P_{\rm rad}$ vary weakly within the star. We can assume $\beta = P_{\rm gas}/P$ to be constant and so $(1-\beta) = P_{\rm rad} / P$. 

All three assumptions of the Eddington stellar model are satisfied for hydrogen-burning main-sequence stars which are hotter than 30\,000\,K. We conclude that they are a reasonable representation of the physical properties of those massive and very massive stars.

\subsubsection{Comparison of the Eddington mass to stellar structure calculations\label{s:comp_mass}}
\begin{figure}
\begin{center}
\resizebox{\hsize}{!}{\includegraphics{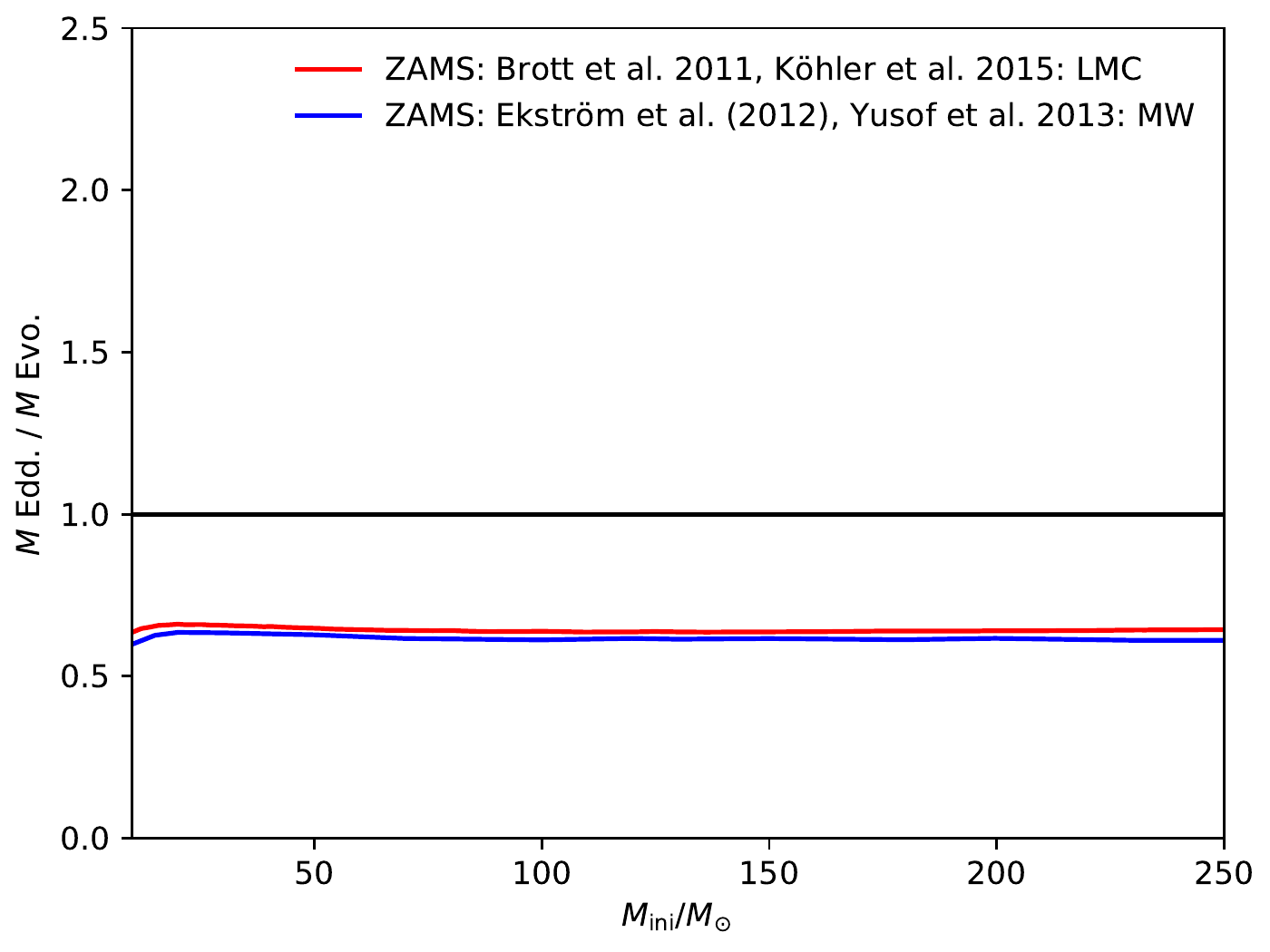}}
\end{center}
\caption{Zero-age main-sequence, initial evolutionary mass versus Eddington mass over current evolutionary mass: The initial condition for stellar evolutionary calculations is a chemical homogeneous star. The ratio between the Eddington mass and evolutionary mass is constant for all stellar masses.}
\label{f:massZAMS}
\end{figure}
\begin{figure}
\begin{center}
\resizebox{\hsize}{!}{\includegraphics{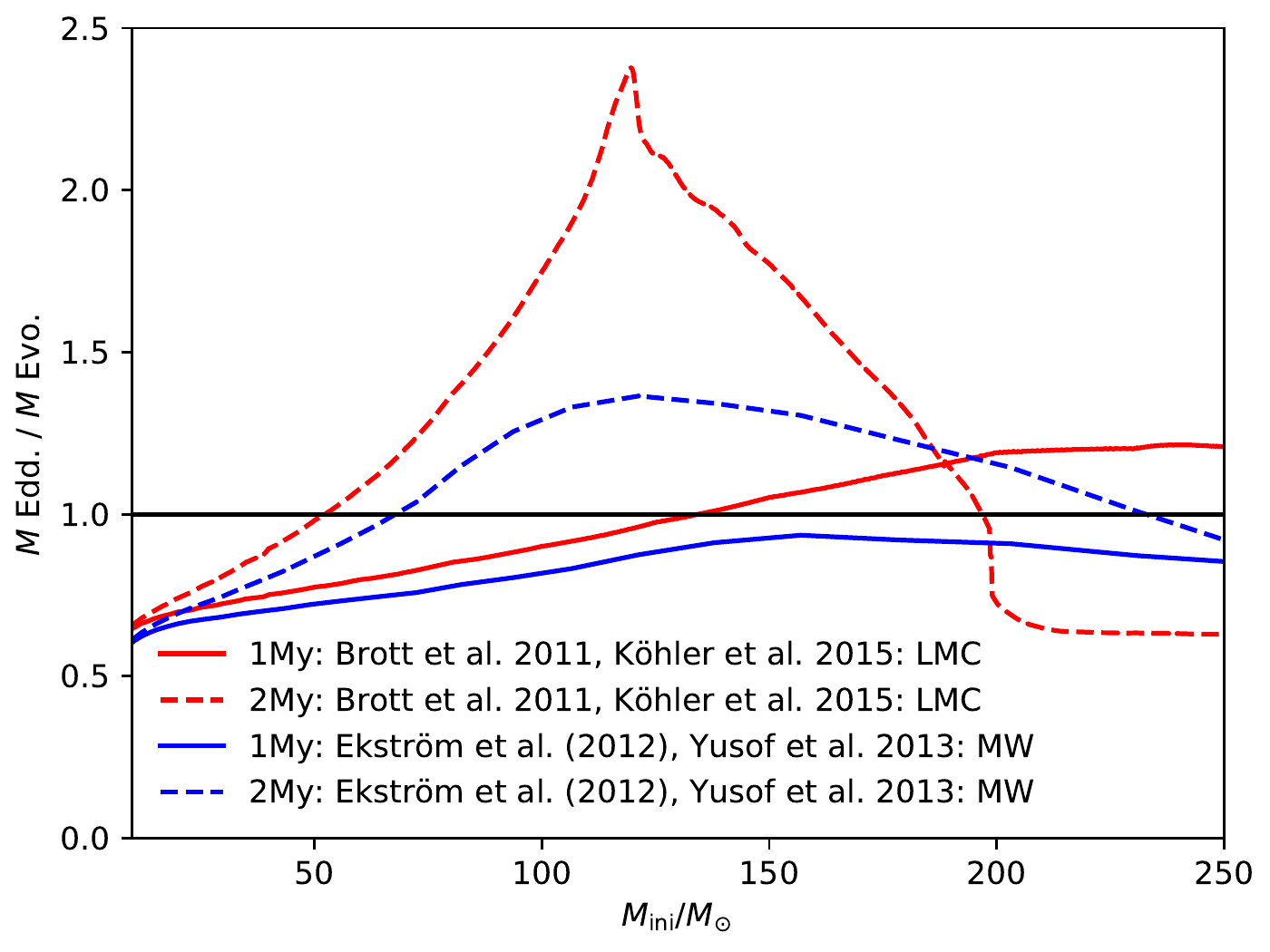}}
\end{center}
\caption{1 and 2 Myr non-rotating main-sequence, initial evolutionary mass versus Eddington mass over current evolutionary mass: The majority of our targets are in the age range between 1 and 2 Myr. The Eddington and evolutionary masses agree within 0.15\,dex. In the mass range between 120 and 130\,$M_{\odot}$ the discrepancy can exceed 0.3\,dex for the \citep{brott2011, koehler2015} tracks.
}
\label{f:mass1-2Myr}
\end{figure}
\begin{figure}
\begin{center}
\resizebox{\hsize}{!}{\includegraphics{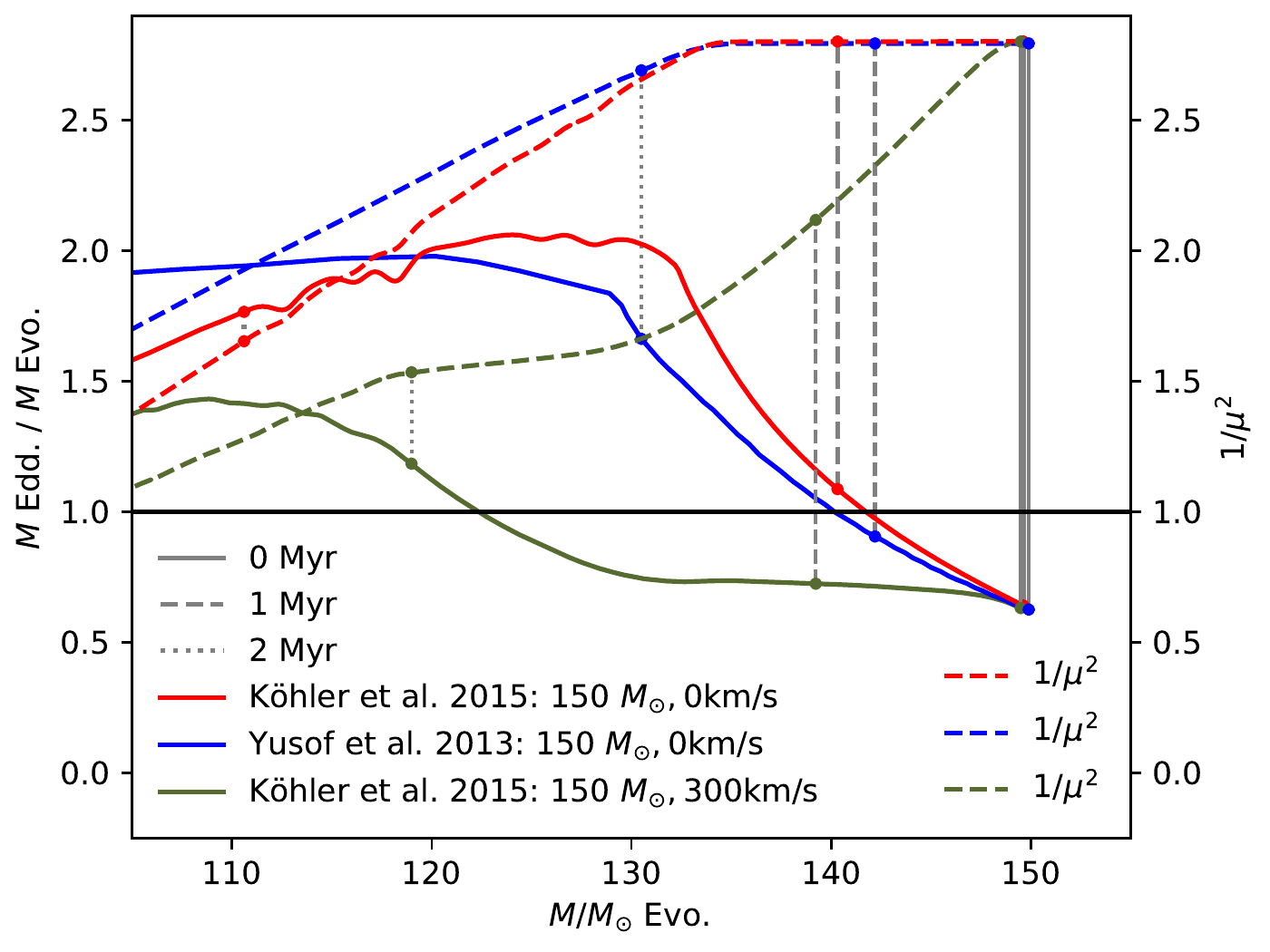}}
\end{center}
\caption{Evolutionary track for a 150\,$M_{\odot}$ star and its mean molecular weight at the stellar surface from \citep{koehler2015} and \citep{yusof2013} for LMC metallicity with an initial rotational velocity of 0 and 300km/s. Grey vertical lines indicate the 0, 1 and 2 Myr time steps.
}
\label{f:150Mtrack}
\end{figure}
To quantify how well the $M-\Gamma_{\rm e}$ relation works we compare Eddington masses with those from evolutionary non-rotating models by \citet{brott2011} \& \citet[hereafter {\sc bonn}]{koehler2015} at LMC metallicity and \citet{ekstroem2012} \& \citet[hereafter {\sc geneva}]{yusof2013} at solar metallicity. As we consider only the electron opacity in our derived $M-\Gamma_{\rm e}$ relation we expect that the Eddington masses under-predict the stellar masses, because the actual Eddington parameter ($\Gamma$) including the line opacity is larger. The advantage in using $\Gamma_{\rm e}$ is that it is approximately constant throughout the star and can be treated as a stellar parameter.

In Fig.\,\ref{f:massZAMS} we show zero-age main-sequences from the {\sc bonn} and {\sc geneva} tracks. The zero-age main-sequence is similar to the initial condition at the beginning of the evolutionary calculation. The star can be approximated as chemically homogeneous and the ratio between the Eddington and the evolutionary mass can assumed to be constant over the entire mass range. The offset between both masses is $\sim$\,0.15\,dex. 

If the offset is constant, we can apply a correction factor to our Eddington masses, but Fig.\,\ref{f:mass1-2Myr} and \ref{f:150Mtrack} clearly show that this is unfortunately not the case. In Fig.\,\ref{f:mass1-2Myr} we compare 1 and 2 Myr main-sequences from the {\sc bonn} and {\sc geneva} tracks, which represent the age range of stars in R136 in the LMC \citep{crowther2016}. They are visualised in the same way as in Fig.\,\ref{f:massZAMS}. The {\sc geneva} main-sequences agree with the Eddington masses within $\pm0.15$\,dex, but the discrepancy for {\sc bonn} can exceed $+0.3$\,dex in the mass range between 120 and 130\,$M_{\odot}$. By looking at the {\sc bonn} 2 Myr main-sequence it appears that stars with an initial mass more than 200\,$M_{\odot}$ are chemically homogeneous again. 

To better understand the reason for the discrepancy between Eddington and evolutionary masses we compare in Fig.\,\ref{f:150Mtrack} the evolutionary tracks of a 150\,$M_{\odot}$ star. As the star evolves through nucleo-synthesis the $L/M$ ratio increases and the mean molecular weight ($\mu$) in the core increases as well. The Eddington mass also increases, because the chemical composition or mean molecular weight at the stellar surface remains unchanged (non-rotating models). When the star has lost more than 10\% of its initial mass the chemical composition at the surface begins to change. The mean molecular weight at the surface increases, the Eddington mass decreases and the discrepancy becomes smaller again. 

By comparing the two non-rotating 150\,$M_{\odot}$ evolutionary tracks at LMC metallicity we see that after 2 Myr the star on the evolutionary track by \cite{koehler2015} has lost around $40\,M_{\odot}$ while the star on the \cite{yusof2013} track only $20\,M_{\odot}$. Looking at the grey vertical lines of Fig.\,\ref{f:150Mtrack} it seems that the star modelled by \cite{koehler2015} evolves faster than the one by \cite{yusof2013} as a result of the higher mass loss. The higher mass-loss rate of the \cite{koehler2015} model leads to a larger discrepancy between Eddington and evolutionary mass. The two stellar tracks represent non-rotating stars, which means that the chemical mixing is not enhanced. The chemical composition at the stellar surface of both modelled stars changes, when around 10\% of the mass is lost. 

The implemented mixing processes are negligible compared to the mass loss. In Fig.\,\ref{f:150Mtrack} we also show an 150\,$M_{\odot}$ evolutionary track from \cite{koehler2015} with an initial rotational velocity of 300\,km/s. $\mu$ at the stellar surface changes straight away and the Eddington mass only slowly increases. After 1.5\,Myr the star has lost $\sim20\,M_{\odot}$, has spun down to 230\,km/s and the chemical mixing is less efficient. The Eddington mass increases more steeply, but the discrepancy between Eddington and evolutionary stays below 0.15dex.

If the star is chemically homogeneous, a constant correction factor over all stellar masses can be applied and the $M-\Gamma_{\rm e}$ relation is in excellent agreement with predictions from stellar structure modelling. However, stars do not evolve chemically homogeneously. The mean-molecular weight at the stellar surface does not represent the actual $\mu$. The $M-\Gamma_{\rm e}$ relation over predicts the stellar mass and the discrepancy can exceed 0.3\,dex with respect to the evolutionary tracks. Overall the Eddington mass agrees reasonable well with evolutionary models if enhanced chemical mixing is present.

\subsubsection{Comparison of the Eddington mass to observations of the star cluster R136\label{s:comp_obs_mass}}
\begin{figure}
\begin{center}
\resizebox{\hsize}{!}{\includegraphics{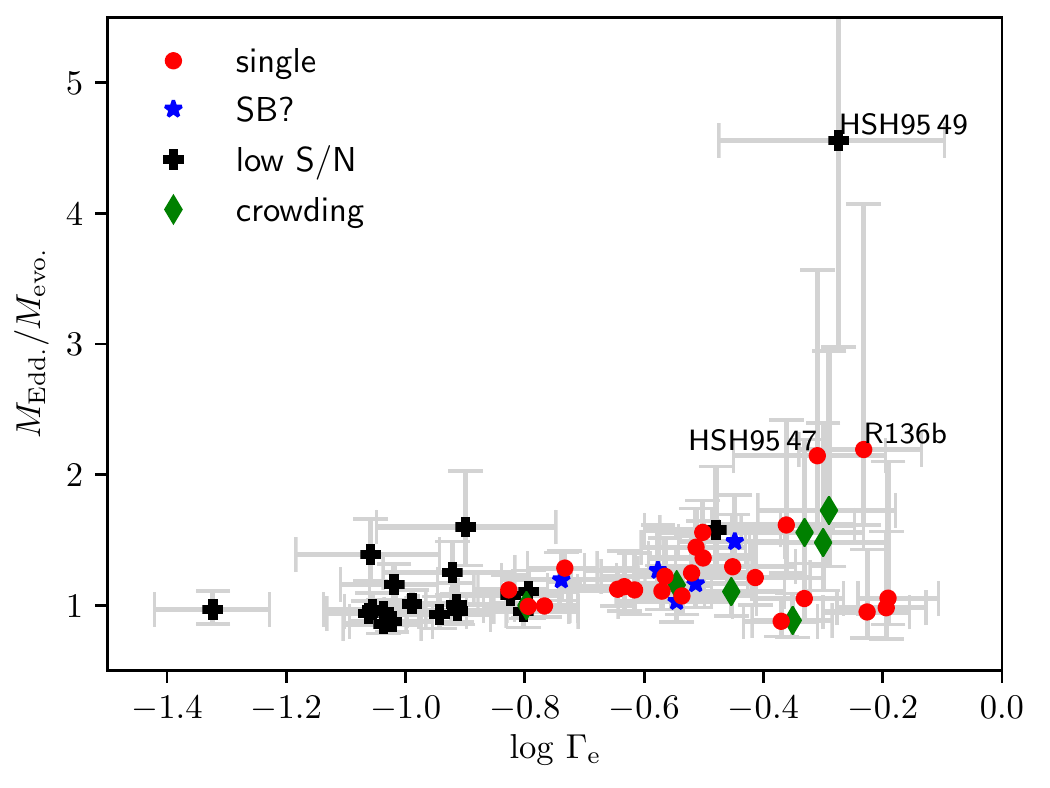}}
\end{center}
\caption{Eddington mass over current evolutionary mass versus $\Gamma_{\rm e}$ for stars in R136 from Bestenlehner et al. (in prep.). Most stars cluster around a constant value except for 3 stars, which have a ratio greater than 2.} 
\label{f:medd-evo_obs}
\end{figure}
The stellar parameters for the stars in R136 are taken from Bestenlehner et al. (in prep.) who performed a spectroscopic analysis with FASTWIND \citep{puls2005} for the O stars and CMFGEN \citep{hillier1998} for the 3 WNh stars using optical spectra taken with STIS on the Hubble Space Telescope \citep{crowther2016}. The stellar masses from Bestenlehner et al. (in prep.) were derived with the BONN Stellar Astrophysics Interface \citep[BONNSAI,][]{schneider2014} using the stellar models from \cite{brott2011, koehler2015}. BONNSAI is a Bayesian tool to calculate the probability distributions of fundamental stellar parameters for a given set of observed stellar parameters including their uncertainties. Spectroscopic masses based on $\log g$ were highly uncertain as the line broadening could not accurately be determined as a result of the low signal-to-noise ratio of the majority of the spectra (Bestenlehner et al. in prep.). Stellar parameters and evolutionary masses from Bestenlehner et al. (in prep.) were used to calculate $\Gamma_{\rm e}$ and the resulting Eddington mass and listed in Table\,\ref{t:mass_gamma}. 

In Fig.\,\ref{f:medd-evo_obs} we compare the Eddington to evolutionary mass ratios to $\Gamma_{\rm e}$. Except for 3 outliers (HSH95 47, HSH95 49 and R136b/HSH95 9) most stars cluster around a constant value. Considering the uncertainties we find an offset of $1.04\pm0.02$. There is an anti-correlation between the Eddington and evolutionary masses. For a given set of stellar parameters smaller evolutionary masses lead to larger Eddington parameters which result in larger Eddington masses and vice versa. Despite the anti-correlation the large discrepancy between Eddington and evolutionary masses as seen for non-rotating stellar models does not occur in our sample. We conclude that our  $M-\Gamma_{\rm e}$ relation works well for this sample as they only show a modest systematic offset.
 
\section{Discussion}
\label{s:disc}
\begin{figure*}
\begin{center}
\resizebox{\hsize}{!}{\includegraphics{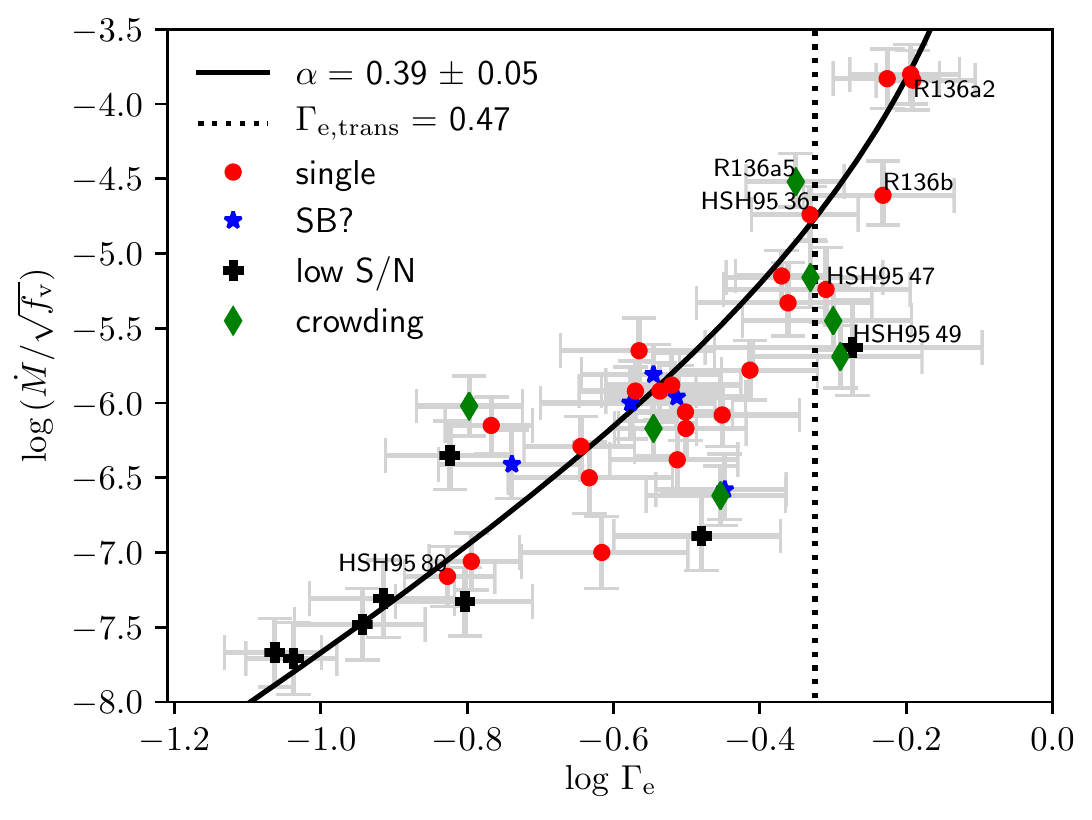}}
\end{center}
\caption{Unclumped $\log \dot{M}$ versus $\log \Gamma_{\rm e}$ for R136 stars from Bestenlehner et al. (in prep.): Black solid line is a fit of the updated CAK-type mass-loss recipe, where the stellar mass is replaced by the Eddington stellar model (Eq.\,\ref{e:cak_mdot-gamma}). Black dotted line indicates the location of the transition Eddington parameter ($\Gamma_{\rm e,trans}$) from optically thin to optically thick winds.} 
\label{f:cak}
\end{figure*}

In this section we verify our new mass-loss recipe. We apply our updated CAK-type mass-loss recipe to stars in the star cluster R136 in the LMC. Eddington parameters are listed in Table\,\ref{t:mass_gamma} and mass-loss rates are from Bestenlehner et al. (in prep.). This is the largest, homogeneously observed data set of stars, which includes terminal velocity measurements from ultraviolet spectra to derive accurate mass-loss rates \citep{crowther2016}. The sample is complete down to $\sim 30\,M_{\odot}$. There are other data sets for early type massive stars available, but with the downside that the terminal velocity to calculated $\dot{M}$ is derived using escape-terminal velocity relations \citep[e.g.][]{lamers1995, kudritzki2000}.

In Fig.\,\ref{f:cak} we compare the unclumped $\dot{M}$ against $\Gamma_{\rm e}$ for R136 O, Of/WN and WNh stars. We used an orthogonal-distance-regression-fitting routine ({\sc odr}) provided by {\it scipy} considering abscissa as well as ordinate errors. Eq.\,\ref{e:cak_mdot-gamma} is a rather complex function to fit. Even though the results are the same we obtain more robust fits by using Eq.\,\ref{e:cak_mdot-gamma} in logarithmic form instead:
\begin{equation}\label{e:log_mdot-gamma}
\log \dot{M} = \log \dot{M}_0 + \left(\frac{1}{\alpha} + 0.5\right) \log(\Gamma_{\rm e}) - \left(\frac{1-\alpha}{\alpha} + 2\right) \log(1-\Gamma_{\rm e})
\end{equation}
with $\dot{M}_0$ including the term which does not contain $\Gamma_{\rm e}$. We derive a value for the force-multiplier parameter $\alpha = 0.39 \pm 0.05$ and present a fit of Eq.\,\ref{e:log_mdot-gamma} through the data in Fig.\,\ref{f:cak}. $\alpha$ is low compared to the expected $\alpha \approx 0.6$ and results in a strong $\Gamma_{\rm e}$ dependency for O stars ($\dot{M} \propto \Gamma_{\rm e}^{3}$). This arises because few O dwarfs possess weak winds.

For an independent test we calculated the other CAK force multiplier parameter $k$ using the derived $\dot{M}$ and $\alpha$ from our fit. We set the thermal velocity to $\varv_{\rm th}=17.4$\,km/s corresponding to a temperature of 45,000\,K for a gas with LMC composition. The electron scattering opacity is estimated using a hydrogen mass-fraction $X = 0.72$ and is $\kappa_{\rm e} \approx 0.34$\,$\mathrm{cm}^2/\mathrm{g}$. The calculated value of $k = 0.14 \pm 0.05$ is reasonable considering that we did not correct $\dot{M}$ for wind clumping or the systematic offset between evolutionary and Eddington masses \citep{pauldrach1986, puls2008}.

\cite{vink2011} explored the high $\Gamma_{\rm e}$-dependent mass-loss behaviour in the transition from optically thin O star winds to optically thick winds of very massive stars. They predicted a sudden change between the two regimes in the form of a ``kink'' at $\Gamma_{\rm e} \sim 0.7$. \cite{bestenlehner2014} observationally confirmed such a ``kink''. In the O star regime they find $\dot{M} \propto \Gamma_{\rm e}^{2.73\pm0.43}$ while in the very massive star regime $\dot{M} \propto \Gamma_{\rm e}^{5.22\pm4.04}$. In the O star regime we find $\dot{M} \propto \Gamma_{\rm e}^{3.06\pm0.28}$ which agrees with \cite{bestenlehner2014} within the uncertainties. Bearing in mind that $\Gamma_{\rm e} \nll 1$ for O stars, but in the range from 0.05 to 0.3, we would expect that the exponent found by \cite{bestenlehner2014} to be greater than ours. However, terminal velocities of the O stars in \cite{bestenlehner2014} were estimated using escape-terminal velocity relations. For the very massive stars we find $\dot{M} \propto 1/(1 - \Gamma_{\rm e})^{3.56\pm0.28}$. $\Gamma_{\rm e,trans}$ is around 0.47. The ``kink'' is at the transition point from an $\dot{M} \propto \Gamma_{\rm e}$ to $\dot{M} \propto 1/(1 - \Gamma_{\rm e})$ dependence. Once the $1/(1-\Gamma_{\rm})$ term dominates ($\Gamma_{\rm e} > \Gamma_{\rm e, trans}$) the $\dot{M}-\Gamma_{\rm e}$ relation becomes very steep and the mass-loss rate is dominated by the Eddington parameter.

In our sample two stars lie close to the transition from optically thin to optically thick winds. HSH95\,36 and R136a5 (HSH95\,20) both have a spectral type of O2 If* and should be still in the optically thin wind regime. Their averaged $\Gamma_{\rm e} = 0.46\pm0.05 \approx \Gamma_{\rm e,trans}$. Interestingly $\Gamma_{\rm e, trans}$ falls into the transition from optically thin to optically thick winds, where the transition mass-loss rate introduced by \cite{vink2012} is also defined. At lower metallicity environments $\Gamma_{\rm e, trans}$ occurs at larger values and vice versa, what is expected as result of the line opacity.

The updated CAK theory reproduces observations which span $30 \leq M \leq 250\,M_{\odot}$, even though the obtained $\alpha$ is relatively small. This is discussed in more detail in the next section (Sect.\,\ref{s:alpha}). The transition from optically thin to optically thick winds occurs at $\Gamma_{\rm e, trans}$, where \cite{vink2012} calibrated the absolute mass-loss rates using the mass-loss rate at this transition. Therefore, we suggest to calibrate the overall mass loss scale of the updated CAK-type mass-loss recipe using the transition mass-loss rate by \cite{vink2012}, if $f_{\rm V}$ is not known, as is usually the case in O stars, or when using the new mass-loss recipe as a mass-loss description for stellar evolution models (Sect.\,\ref{s:mdot_pred}).

\subsection{CAK $\alpha$ parameter}
\label{s:alpha}

The determined $\alpha$ parameter is an effective value for all stars in our sample. The fit includes stars with very weak winds such as OVz dwarfs as well as the strong winds of very massive WNh stars. The way $\alpha$ is defined in CAK, we would not expect a unified $\alpha$ for all stars. With increasing emission line strength ($\Gamma_{\rm e} \rightarrow 1$) $\alpha$ should have lower values, which would lead to an even stronger $1/(1-\Gamma_{\rm e})^x$ dependence with a high exponent $x$. In addition, stars with optically thick winds are generally hydrogen-depleted and therefore the $1/\mu^2$ term in Eq.\,\ref{e:cak_mdot-gamma} changes as well with $\Gamma_{\rm e}$ approaching unity.

\cite{kudritzki1999} introduced the modified wind-momentum ($D_{\rm mom} = \dot{M} \varv_{\infty} \sqrt{R}$), which scales with bolometric luminosity. The modified wind-momentum -- luminosity relation (WLR) has the form
\begin{equation}
\log D_{\rm mom} = \log D_0 + x \log(L/L_{\odot}).
\end{equation}
The inverse of the slope $x$ can be interpreted as an effective $\alpha$ ($\alpha = 1/x$). Bestenlehner et al. (in prep.) find a WLR slope of $x = 2.41\pm 0.13 \rightarrow \alpha = 0.41\pm 0.02$, which is consistent to what we find using the {new mass-loss recipe}. However, \cite{vink2000, vink2001} predicts a shallower WLR slope of $x = 1.83$ corresponding to $\alpha = 0.55$ which is metallicity independent. $\alpha$ is weakly metallicity dependent and decreases with decreasing metallicity \citep{puls2000}. Therefore, the WLR should be steeper at lower metallicity. In the context of the updated CAK-type mass-loss recipe the mass-loss rate depends more strongly on the Eddington parameter in metal-poor than in metal-rich environments. A steeper dependence on $\Gamma_{\rm e}$ for more metal poor environments was recently found also for hydrogen-depleted {\it classical} WR stars \citep{sander2020}.

The updated CAK-wind theory explains the observed mass-loss dependence on the Eddington parameter. $\Gamma_{\rm e}$ is approximately independent of the radius and can be treated as a stellar parameter like luminosity or effective temperature in stellar structure calculations. In addition, $\Gamma_{\rm e} \propto T_{\rm eff}^4/g = \mathcal{L}$ the inverse flux weighted gravity defined as the spectroscopic luminosity $\mathcal{L}$ can be used instead \citep{langer2014}, if the distance or extinction to the star is not known or highly uncertain. This only applies to O stars with optically thin winds as in the optically thick wind regime $\log g$ cannot be constrained.

\subsection{Mass loss prediction for stellar evolutionary models}
\label{s:mdot_pred}
\begin{table*}
	\centering
	\caption{Observed $\log\dot{M}~(M_{\odot}\mathrm{yr}^{-1})$ from Bestenlehner et al. (in prep.) and predicted $\dot{M}$ using \citet{vink2000, vink2001} derived with BONNSAI \citep{schneider2014}, mass-loss recipe for hydrogen-rich WNL stars by \citet{graefener2008} and with the new mass-loss recipe for 2 representative stars. Observed and updated CAK-type $\dot{M}$ are corrected for an volume filling factor $f_{\rm V} = 0.1$.}
	\label{t:mdot}
	\begin{tabular}{lccccc}
		\hline
		Star & SpT & observed & \citet{vink2000, vink2001} & new $\dot{M}$ recipe & \citet{graefener2008}\\
		\hline
       	R136a2/HSH95 5 & WN5h & $-4.34\pm 0.20$ & $-4.45$ & $-4.24$ & $-4.86$\\
		HSH95 80       & O8V  &  $-7.66\pm0.20$  & $-6.89$ & $-7.55$ & --      \\
		\hline
	\end{tabular}
\end{table*}

The new mass-loss recipe can be readily implemented as a mass-loss description in stellar evolutionary calculations of main-sequence massive stars. It not only does match the mass-loss rates of O stars but also the enhanced mass-loss rates of WNh stars. In principle our updated CAK-type mass-loss recipe might be also applicable for {\it classical} WR stars of spectral type WN, WC and WO. However, in hydrogen deficient WR stars the electron scattering opacity is less dominant and bound-free and/or line opacities might need to be considered as well. This will be the topic for a future study. In this section we compare observed with predicted mass-loss rates for the {\sc bonn} models \citep{vink2000, vink2001}, \cite {graefener2008} for hydrogen-rich late WN stars (WNL, $T_{\star} \lessapprox 70\,000$\,K at optical depth $\tau =20$) and our new mass-loss recipe. At the end of this section we outline how the updated CAK-type mass-loss recipe could be implemented into stellar structure calculations.

Observed mass-loss rates for R136 stars are taken from Bestenlehner et al. (in prep.). We assumed a typical volume filling factor $f_{\rm V} = 0.1$ for O and WNh stars and scale the mass-loss rates accordingly, which is justified by the electron scattering wings of the WNh stars. BONNSAI \citep{schneider2014} and the stellar parameters from Bestenlehner et al. (in prep.) were combined to have mass-loss rate predictions based on the standard mass-loss recipes \citep{vink2000, vink2001} implemented into the {\sc bonn} tracks \citep{brott2011, koehler2015}. Updated CAK-type mass-loss rates were calculated using the fit shown in Fig.\,\ref{f:cak} and scaled down for $f_{\rm V} = 0.1$. In addition, we computed $\dot{M}$ using the mass-loss recipe by \cite{graefener2008} for WNh stars, which is implemented into the {\sc geneva} stellar evolution code \citep{yusof2013}. Stellar parameters were taken from Bestenlehner et al. (in prep.) and a metallicity of $Z/Z_{\odot}=0.5$ was assumed for the LMC. We chose the apparent single stars R136a2 (HSH95\,5, WN5h) with the highest $\Gamma_{\rm e} = 0.64$ and HSH95\,80 (O8V) with the lowest $\Gamma_{\rm e} = 0.15$ in the sample of Bestenlehner et al. (in prep.). They cover a $\dot{M}$ range of more than 3\,dex.

In Table\,\ref{t:mdot} we summarise the different mass-loss rates for comparison purposes. Observed and updated CAK $\dot{M}$ well agree within the uncertainties as the new mass-loss recipe is a fit through these data. There is an offset of $0.1$\,dex, because both stars lie by chance below the updated CAK-type mass-loss rate fit. Stellar evolutionary mass-loss rates based on \citep{vink2000, vink2001} are slightly lower for the WNh star R136a2 ($0.11$\,dex), but are 0.77\,dex too high for the O8 dwarf. This suggests that mass-loss rates for O dwarfs are over-predicted in stellar structure calculation. Mass-loss predictions based on \cite{graefener2008} are $\sim 0.5$\,dex lower for R136a2. However, the {\sc geneva} evolutionary code uses \cite{vink2000, vink2001} recipe if the predicted mass-loss rate by \cite{graefener2008} is smaller than this \citep{yusof2013}.

Before $\dot{M}$ predictions based on the updated CAK-type mass-loss recipe can be implemented we need to find a typical value for the force multiplier parameter $\alpha$ and calibrate the absolute mass-loss rate scale for a range of metallicities. To derive accurate stellar wind parameters ultraviolet observations are necessary. The director's discretionary program Hubble UV Legacy Library of Young Stars as Essential Standards (ULLYSES)\footnote{\url{http://www.stsci.edu/stsci-research/research-topics-and-programs/ullyses}} with HST will provide an ultraviolet spectroscopic library of hot stars over a wide range of metal-poor environments. Once the stellar and wind parameters are derived the new mass-loss rate recipe is used to determine $\alpha$ and the absolute mass-loss rate scale by fitting Eq.\,\ref{e:log_mdot-gamma} through the data. In the absence of objects with optically thick winds the WLR can be used instead. The transition mass-loss rate ($\dot{M}_{\rm trans}$), introduced by \cite{vink2012}, can be applied to calibrate the absolute $\dot{M}$ scale ($\log \dot{M}_0$, Eq.\,\ref{e:log_mdot-gamma}) and to determine an effective volume filling factor. $\dot{M}_{\rm trans}$ by \cite{vink2012} is defined at the transition from optically thin to optically thick winds at a unique point where the wind efficiency is equal the optical depth at the sonic point equal to unity ($\eta = \tau = 1$). This is the same point at which our $\Gamma_{\rm e, trans}$ is defined. 

Using Eq.\,\ref{e:m-gamma} with values of $\Gamma_{\rm e, trans} = 0.47\pm0.02$ we find a transition luminosity $\log L_{\rm trans}/L_{\odot} \approx 6.35\pm0.01$ which is consistent with bolometric luminosity of R136a5 and HSH95\,36. $\log \dot{M}_{\rm trans} = -4.76\pm 0.18$ at $\Gamma_{\rm e, trans} = 0.47$. Using Eq.\,12 from \cite{vink2012} with our $L_{\rm trans}$ and $\varv_{\infty} \approx 3300$\,km/s based on R136a5 and HSH95\,36 we find $\log \dot{M}_{\rm trans} = -5.08\pm0.04$. This corresponds to $f_{\rm V} = 0.23_{-0.15}^{+0.40}$ or a clumping factor $D~(=1/f_{\rm V})$ in the range of 2 to 12. $f_{\rm V}$ is larger than the value assumed above for the comparison ($f_{\rm V} = 0.1$), but it still falls within the uncertainty interval. A larger volume filling factor would suggest that the mass-loss rate of the very massive WNh stars are underestimated by about a factor of 2 in the {\sc bonn} models. However, this is only an indication and a larger sample is required which will be provided by ULLYSES in combination with optical ground based observations such as the 4MOST/1001MC survey for the Magellanic Clouds \citep{cioni2019}.

Nevertheless we conclude that the updated CAK-type mass-loss recipe reproduces the observations and the overall mass-loss rate scale is in line with our current understanding of the stellar winds of massive and very massive stars.

\section{Conclusion}
\label{s:con}

The new mass-loss recipe is a neat extension to the mass-loss formula by CAK. It combines the optically thin wind regime of O stars with the optically thick wind regime of very massive WNh stars. The transition occurs at $\Gamma_{\rm e, trans}$ where $\Gamma_{\rm e}$ dependence at the O star regime turns into a $1/(1-\Gamma_{\rm e})$ dependence for the enhanced mass loss of WNh stars. The updated CAK-type mass-loss recipe keeps the simplicity of the original CAK wind theory, which made CAK so widely used. It only requires the force multiplier parameter $\alpha$, mean molecular weight and absolute mass-loss rate scale for given metallicity. The simplicity and universal approach of the new mass-loss recipe makes it suitable to be used as a mass-loss description in stellar structure calculations for massive main-sequence stars with $T_{\rm eff} \geq 30\,000$\,K, but might be able to be applied to massive post-main sequence stars as well. A future study will explore the validity of this wind theory for {\it classical} hydrogen free WR stars and hydrogen stripped stars.

The CAK parameter $\alpha$ varies with metallicity. Once $\alpha$ is known we know the $\Gamma_{\rm e, trans}$ and are able to calibrate the mass-loss predictions for a given metallicity environment with the method outlined in \cite{vink2012}. This is in particular of interest for very massive stars at low metallicty in the early and high redshift universe. Very massive stars play a key role in the re-ionisation of the young universe and dominate the strong He\,{\sc ii}\,$\lambda1640$ emission in the ultraviolet \citep{crowther2019}. State of the art population synthesis models such as {\sc starburst99} \citep{levesque2013} and {\sc bpass} \citep{eldridge2017} are not able to predict the required emission line strength. Our wind theory predicts that very massive stars at low metallicity should also have optically thick winds, if their $\Gamma_{\rm e}$ is greater than $\Gamma_{\rm e, trans}$. The inclusion of very massive stars with an increased $\dot{M}$ can leverage population synthesis models in reproducing emission lines in the ultraviolet.

\section*{Acknowledgements}
I thank the anonymous referee for providing constructive comments and suggestions which improved the clarity and content of the manuscript. JMB acknowledges financial support from the University of Sheffield. I would like to thank Joachim Puls for fruitful discussions and Paul Crowther and Jorick Vink for constructive comments to this manuscript. The paper is dedicated to my late mother Liselotte.

%%%%%%%%%%%%%%%%%%%%%%%%%%%%%%%%%%%%%%%%%%%%%%%%%%



\bibliographystyle{mnras}
\bibliography{extracted} % if your bibtex file is called example.bib

%%%%%%%%%%%%%%%%%%%%%%%%%%%%%%%%%%%%%%%%%%%%%%%%%%

%%%%%%%%%%%%%%%%% APPENDICES %%%%%%%%%%%%%%%%%%%%%
\appendix
\onecolumn
\section{Eddington parameters, evolutionary and Eddington masses}

\begin{table*}
	\centering
	\caption{Eddington parameters were computed with stellar parameters and evolutionary masses from Bestenlehner et al. (in prep.). Eddington masses were calculated using Eq.\,\ref{e:m-gamma} and those Eddington parameters.}
	\label{t:mass_gamma}
	\begin{tabular}{lccc}
		\hline
		Star & $\Gamma_{\rm e}$ & evolutionary masses ($M_{\rm evo.}/M_{\odot}$) & Eddington mass ($M_{\rm Edd.}/M_{\odot}$)\\
		\hline
R136a1     & $0.64^{+0.11}_{-0.11}$ & $ 214.8^{+  45.2}_{-  30.5}$ & $ 210.5^{+ 246.0}_{ -98.8}$\\
R136a2     & $0.64^{+0.14}_{-0.07}$ & $ 187.2^{+  23.0}_{-  33.3}$ & $ 197.2^{+ 390.3}_{ -67.5}$\\
R136a3     & $0.59^{+0.11}_{-0.09}$ & $ 153.6^{+  28.4}_{-  23.3}$ & $ 145.6^{+ 144.2}_{ -57.0}$\\
R136a4     & $0.51^{+0.15}_{-0.12}$ & $  86.2^{+  27.2}_{-  19.5}$ & $ 148.7^{+ 204.8}_{ -66.1}$\\
\medskip
R136a5     & $0.45^{+0.07}_{-0.06}$ & $ 105.2^{+  17.9}_{-  14.8}$ & $  92.9^{+  40.2}_{ -24.0}$\\
R136a6     & $0.43^{+0.06}_{-0.06}$ & $ 111.6^{+  17.5}_{-  14.6}$ & $  97.8^{+  35.1}_{ -22.7}$\\
R136a7     & $0.50^{+0.14}_{-0.12}$ & $  87.8^{+  28.9}_{-  19.2}$ & $ 129.9^{+ 155.1}_{ -57.7}$\\
R136b      & $0.59^{+0.15}_{-0.13}$ & $  93.2^{+  26.5}_{-  18.7}$ & $ 204.2^{+ 345.4}_{ -99.9}$\\
HSH95\,30  & $0.32^{+0.05}_{-0.05}$ & $  39.6^{+   7.1}_{-   5.4}$ & $  61.7^{+  15.6}_{ -12.1}$\\
\medskip
HSH95\,31  & $0.39^{+0.09}_{-0.08}$ & $  67.0^{+  16.7}_{-  12.8}$ & $  81.1^{+  43.2}_{ -23.8}$\\
HSH95\,35  & $0.30^{+0.07}_{-0.06}$ & $  46.6^{+  10.7}_{-   9.1}$ & $  58.1^{+  22.7}_{ -13.3}$\\
HSH95\,36  & $0.47^{+0.08}_{-0.08}$ & $ 117.6^{+  23.7}_{-  16.5}$ & $ 123.6^{+  57.9}_{ -37.8}$\\
HSH95\,40  & $0.35^{+0.10}_{-0.07}$ & $  54.2^{+  13.5}_{-  11.7}$ & $  70.1^{+  39.6}_{ -19.1}$\\
HSH95\,45  & $0.36^{+0.08}_{-0.07}$ & $  50.0^{+  12.1}_{-   8.8}$ & $  74.3^{+  31.0}_{ -20.0}$\\
\medskip
HSH95\,46  & $0.47^{+0.12}_{-0.11}$ & $  79.6^{+  24.2}_{-  16.2}$ & $ 123.9^{+ 106.9}_{ -49.1}$\\
HSH95\,47  & $0.49^{+0.15}_{-0.14}$ & $  64.8^{+  24.7}_{-  15.1}$ & $ 138.9^{+ 176.5}_{ -65.1}$\\
HSH95\,48  & $0.44^{+0.13}_{-0.11}$ & $  66.0^{+  22.0}_{-  15.3}$ & $ 106.5^{+  99.9}_{ -41.6}$\\
HSH95\,49  & $0.53^{+0.27}_{-0.20}$ & $  37.8^{+  22.3}_{-  12.7}$ & $ 172.3^{+ 993.5}_{-104.8}$\\
HSH95\,50  & $0.28^{+0.04}_{-0.03}$ & $  46.6^{+   6.1}_{-   5.9}$ & $  53.8^{+  11.0}_{  -7.5}$\\
\medskip
HSH95\,52  & $0.27^{+0.06}_{-0.04}$ & $  45.2^{+   8.8}_{-   7.8}$ & $  50.1^{+  14.6}_{  -9.3}$\\
HSH95\,55  & $0.29^{+0.06}_{-0.05}$ & $  51.6^{+  10.4}_{-   8.9}$ & $  55.2^{+  17.3}_{ -11.1}$\\
HSH95\,58  & $0.35^{+0.08}_{-0.07}$ & $  63.0^{+  16.6}_{-  11.7}$ & $  69.5^{+  30.7}_{ -19.6}$\\
HSH95\,62  & $0.28^{+0.07}_{-0.06}$ & $  50.0^{+  12.7}_{-   9.7}$ & $  51.4^{+  18.6}_{ -12.1}$\\
HSH95\,64  & $0.31^{+0.06}_{-0.06}$ & $  41.2^{+   9.7}_{-   7.2}$ & $  59.5^{+  20.1}_{ -14.0}$\\
\medskip
HSH95\,65  & $0.32^{+0.07}_{-0.06}$ & $  45.4^{+  10.7}_{-   7.8}$ & $  61.8^{+  21.4}_{ -14.8}$\\
HSH95\,66  & $0.27^{+0.07}_{-0.06}$ & $  41.6^{+  11.6}_{-   8.8}$ & $  50.8^{+  19.8}_{ -12.4}$\\
HSH95\,68  & $0.33^{+0.09}_{-0.08}$ & $  42.2^{+  13.4}_{-   9.3}$ & $  66.5^{+  35.2}_{ -20.3}$\\
HSH95\,69  & $0.23^{+0.04}_{-0.04}$ & $  36.6^{+   7.2}_{-   5.7}$ & $  41.1^{+   8.8}_{  -6.9}$\\
HSH95\,70  & $0.31^{+0.07}_{-0.06}$ & $  51.0^{+  12.8}_{-   9.9}$ & $  59.3^{+  23.6}_{ -14.5}$\\
\medskip
HSH95\,71  & $0.24^{+0.08}_{-0.06}$ & $  37.8^{+  11.2}_{-   9.0}$ & $  42.2^{+  17.4}_{ -10.0}$\\
HSH95\,73  & $0.18^{+0.02}_{-0.02}$ & $  26.0^{+   4.0}_{-   3.0}$ & $  33.4^{+   4.4}_{  -4.1}$\\
HSH95\,75  & $0.18^{+0.04}_{-0.04}$ & $  27.6^{+   7.2}_{-   5.3}$ & $  32.9^{+   7.9}_{  -6.1}$\\
HSH95\,78  & $0.26^{+0.08}_{-0.07}$ & $  38.8^{+  12.7}_{-   9.1}$ & $  49.1^{+  21.7}_{ -13.1}$\\
HSH95\,80  & $0.15^{+0.02}_{-0.02}$ & $  24.6^{+   3.5}_{-   3.4}$ & $  27.5^{+   3.9}_{  -2.9}$\\
\medskip
HSH95\,86  & $0.16^{+0.03}_{-0.02}$ & $  29.2^{+   5.3}_{-   4.5}$ & $  29.1^{+   4.9}_{  -3.8}$\\
HSH95\,90  & $0.17^{+0.02}_{-0.02}$ & $  31.2^{+   4.9}_{-   3.8}$ & $  31.0^{+   4.0}_{  -3.7}$\\
HSH95\,92  & $0.16^{+0.03}_{-0.02}$ & $  29.6^{+   4.2}_{-   4.3}$ & $  29.3^{+   4.6}_{  -3.1}$\\
HSH95\,94  & $0.23^{+0.07}_{-0.05}$ & $  37.0^{+  10.8}_{-   8.6}$ & $  42.2^{+  16.1}_{  -9.7}$\\
HSH95\,108 & $0.12^{+0.03}_{-0.03}$ & $  23.2^{+   6.1}_{-   4.7}$ & $  23.3^{+   4.7}_{  -3.7}$\\
\medskip
HSH95\,112 & $0.16^{+0.03}_{-0.03}$ & $  25.4^{+   5.5}_{-   4.3}$ & $  28.1^{+   5.3}_{  -4.3}$\\
HSH95\,114 & $0.16^{+0.04}_{-0.03}$ & $  29.0^{+   7.1}_{-   5.5}$ & $  27.5^{+   6.0}_{  -4.6}$\\
HSH95\,116 & $0.09^{+0.01}_{-0.01}$ & $  19.0^{+   3.1}_{-   2.4}$ & $  18.1^{+   1.9}_{  -1.8}$\\
HSH95\,120 & $0.09^{+0.02}_{-0.02}$ & $  19.0^{+   4.0}_{-   3.5}$ & $  18.4^{+   2.9}_{  -2.2}$\\
HSH95\,121 & $0.09^{+0.01}_{-0.01}$ & $  19.6^{+   3.3}_{-   2.6}$ & $  16.8^{+   1.8}_{  -1.7}$\\
\medskip
HSH95\,123 & $0.11^{+0.02}_{-0.02}$ & $  22.8^{+   5.4}_{-   4.1}$ & $  21.2^{+   3.6}_{  -3.0}$\\
HSH95\,129 & $0.05^{+0.01}_{-0.01}$ & $  12.8^{+   3.2}_{-   2.5}$ & $  12.4^{+   1.8}_{  -1.5}$\\
HSH95\,132 & $0.12^{+0.03}_{-0.02}$ & $  23.4^{+   5.0}_{-   4.0}$ & $  22.4^{+   3.7}_{  -3.0}$\\
HSH95\,134 & $0.09^{+0.01}_{-0.01}$ & $  19.4^{+   3.4}_{-   2.6}$ & $  18.2^{+   2.0}_{  -1.8}$\\
HSH95\,135 & $0.10^{+0.02}_{-0.02}$ & $  19.4^{+   3.5}_{-   2.7}$ & $  19.6^{+   2.3}_{  -2.2}$\\
\medskip
HSH95\,139 & $0.09^{+0.02}_{-0.01}$ & $  21.2^{+   3.8}_{-   3.1}$ & $  18.5^{+   2.2}_{  -2.0}$\\
HSH95\,141 & $0.10^{+0.02}_{-0.02}$ & $  16.8^{+   3.9}_{-   2.6}$ & $  19.5^{+   2.5}_{  -2.6}$\\
HSH95\,143 & $0.15^{+0.03}_{-0.03}$ & $  25.8^{+   5.8}_{-   4.5}$ & $  27.6^{+   5.2}_{  -4.2}$\\
HSH95\,159 & $0.12^{+0.03}_{-0.03}$ & $  18.4^{+   5.6}_{-   4.0}$ & $  23.0^{+   5.0}_{  -4.1}$\\
HSH95\,162 & $0.13^{+0.05}_{-0.04}$ & $  15.0^{+   6.2}_{-   4.4}$ & $  24.0^{+   8.3}_{  -5.4}$\\
%\medskip
HSH95\,173 & $0.09^{+0.03}_{-0.02}$ & $  13.2^{+   4.4}_{-   3.1}$ & $  18.3^{+   3.9}_{  -3.2}$\\
	\hline
	\end{tabular}
\end{table*}

%%%%%%%%%%%%%%%%%%%%%%%%%%%%%%%%%%%%%%%%%%%%%%%%%%

% Don't change these lines
\bsp	% typesetting comment
\label{lastpage}
\end{document}